
\input phyzzx
\FRONTPAGE
\line{\hfill BROWN-HET-957}
\line{\hfill Revised version}
\line{\hfill July 1994}
\bigskip
\titlestyle{{\bf UNIVERSE REHEATING AFTER INFLATION}\foot{Work
supported in part by the Department of Energy under contract
DE-FG02-91ER40688 - Task A}\break}
\author {Y. Shtanov,$^{(1,2)}$ J. Traschen$^{(3)}$ and R.
Brandenberger$^{(2)}$}
\vskip.25in
\item {1)} {\it Bogolyubov Institute for Theoretical Physics, Kiev, 252143,
Ukraine}
\item {2)} {\it Department of Physics, Brown University, Providence, Rhode
Island, 02912, USA}
\item {3)} {\it Department of Physics \& Astronomy, University of
Massachusetts, Amherst, Massachusetts, 01003, USA}
\endpage
\abstract

We study the problem of scalar particle production after
inflation by an inflaton field which is oscillating rapidly relative to the
expansion of the universe. We use the framework of the
chaotic inflation scenario with quartic and quadratic inflaton potentials.
Particles produced are described by a quantum scalar field $\chi$, which
is coupled to the inflaton via linear and quadratic couplings.
The particle production effect is studied
using the standard technique of Bogolyubov transformations.
Particular attention
is paid to parametric resonance phenomena which take place in the
presence of the quickly oscillating inflaton field.
We have found that in the region of applicability of perturbation theory
the effects of parametric resonance are crucial, and estimates based on
first order Born approximation often underestimate the particle production.
In the case of the quartic
inflaton potential $V(\varphi) = \lambda \varphi^4$, the particle production
process is very efficient for either type of
coupling between the inflaton field and the scalar field $\chi$ even for
small values of coupling constants. The reheating temperature of the universe
in this case is
$\left[\lambda\, \log\, (1 / \lambda)\right]^{- 1}$ times larger than the
corresponding estimates based on first order Born approximation.
In the case of the quadratic inflaton potential the reheating process
depends crucially on the type of coupling between the inflaton and the scalar
field $\chi$ and on the magnitudes of the coupling constants.
If the inflaton coupling to fermions and its linear (in inflaton field)
coupling to scalar fields are suppressed, then, as previously discussed by
Kofman, Linde and Starobinsky (see e.g. Ref. 13), the
inflaton field will eventually decouple from the rest of the matter,
and the residual inflaton
oscillations may provide the (cold) dark matter of the universe.
In the case of the quadratic inflaton potential we obtain the lowest and the
highest possible
bounds on the effective energy density of the inflaton field when it freezes
out.

\endpage

{\bf \chapter{Introduction}}

According to the simplest version of the inflationary scenario (see
\REF\l{A.Linde,
{\it Particle Physics and Inflationary Cosmology}
(Harwood, Chur, Switzerland, 1990).} [\l]) the
universe in the past expands almost exponentially
with time (such an expansion is called ``inflation'') while its energy density
is dominated by the effective
potential energy density of a special scalar field $\varphi$, called
the inflaton.
Sooner or later inflation terminates and the inflaton field starts
quasiperiodic motion with slowly decreasing amplitude. Right after inflation
the universe is empty of particles, i.e. ``cold''. Quasiperiodic
evolution of the
inflaton field leads to creation of particles of various kinds, after
thermalization of which due to collisions and decays the universe becomes
``hot''. The role of the inflaton field in the inflationary scenario is not
necessarily played by a fundamental scalar field. It can be the expectation
value of a scalar operator constructed of fields of other type. It can also
be the effective scalar introduced in the inflationary models
based on a higher derivative theory of gravity. \Ref\s{A.Starobinsky,
{\it Phys. Lett.} {\bf 91B}, 99 (1980).} Almost all such scenarios have
an important feature: the inflationary stage
is followed by quasiperiodic evolution of the effective inflaton field
that leads to particle creation.

In this work we are going to study the transition of the universe from
an inflationary
to a hot stage sketched above. The problem is not new, and there are
many papers devoted to its study using different methods in the context of
various inflationary scenarios \REFS\dl{A.Dolgov and A.Linde, {\it Phys. Lett.}
{\bf 116B}, 329 (1982).} \REFSCON\afw{L.Abbott, E.Farhi and M.Wise,
{\it Phys. Lett.} {\bf 117B}, 29 (1982).} \REFSCON\astw{A.Albrecht,
P.Steinhardt, M.Turner and F.Wilczek, {\it Phys. Rev. Lett.} {\bf 48}, 1437
(1982).} \REFSCON\nos{D.Nanopoulos, K.Olive and M.Srednicki, {\it Phys. Lett.}
{\bf 127B}, 30 (1983).} \REFSCON\ms{M.Morikawa and M.Sasaki, {\it Progr.
Theor. Phys.} {\bf 72}, 782 (1984).} \REFSCON\hs{A.Hosoya and M.Sakagami,
{\it Phys. Rev.} {\bf D29}, 2228 (1984).} \refsend\ (see also Refs. \l\ and
\REF\kt{E.Kolb and M.Turner, {\it The Early Universe} (Addison-Wesley
Publishing Company, Redwood City, California, 1990).} \kt).
There is nevertheless
a drawback common to most of them - the calculations were
based on ordinary perturbation theory. The rates of particle
production by the oscillating inflaton field were
calculated in first order Born approximation. However, as it has been shown
in  \REF\tb{J.Traschen and R.Brandenberger, {\it Phys. Rev.} {\bf D42}, 2491
(1990).} Ref. \tb\
this approach disregards possible parametric resonance effects which can
enhance the rate of boson particle production. To detect such effects one has
to go beyond standard perturbation theory. Such resonance amplification
effects have been studied in [\tb] in the context of the new inflationary
scenario.
It was shown that typically the oscillating inflaton field produces
many particles due to
parametric resonance effect even in the case of extremely small couplings,
and that estimates based on ordinary
perturbation theory miss the largest effect. Parametric resonance
effects as relevant to reheating of the universe were also studied in
\REF\dk{A.Dolgov and D.Kirilova, {\it Sov. Nucl. Phys.} {\bf 51}, 273 (1990).}
Ref. \dk\ where it was also demonstrated that there is no
resonance enhancement of fermionic particle production.

In the present paper we are going to study the problem in a systematic
way. We will consider a model with inflation based on a power law potential
for the inflaton field. The inflaton field $\varphi$ throughout this paper
will be regarded as classical. In Section 2 we will describe the
evolution
of the inflaton field after inflation without taking into account its
couplings to other fields. Couplings of the inflaton field $\varphi$ to
fields describing fermionic spin-$1/2$ and bosonic scalar particles
will be considered in the following sections. In particular, the
process of production of these particles by the oscillating inflaton field
will be studied. In all the cases we will assume the
couplings to be sufficiently small so that perturbation theory is valid.
In Section 3 we will derive the expressions for particle
production rates in first order Born approximation of ordinary
perturbation theory. The material of Sections 2 and 3 will closely follow a
previous work \REF\sh{Y.Shtanov, {\it Ukr. Fiz. Zh.} {\bf 38}, 1425 (1993)
(in Russian).} [\sh]. In Section 4 we study general aspects of the parametric
resonance effect, and in Sections 5 and 6 we apply the results obtained
to the
concrete cases of quartic and quadratic inflaton potentials. In Section 7
we present our general conclusions.

It will be shown that in many important cases it is not legitimate to neglect
parametric resonance effects. These effects lead to a great enhancement
of the particle production as compared to the estimates based on ordinary
perturbation theory. The number of particles $N_k$ in a mode with wavenumber
$k$ that has experienced the resonance turns out to be much larger than unity.
The universe reheating temperature therefore can turn out to be
sufficiently large even for small values of the couplings between inflaton
and other fields.

The opinion of the authors of Ref. \dk\ is that parametric resonance effects
of scalar particle production
are likely to be suppressed by the processes of scattering and decay of the
particles produced. The reason is that particle scattering
and decay will tend to reduce the mean occupation numbers $N_k$ in each
particular mode so that these numbers will grow not so fast as
to be able to become large during the resonance period. This question
is very important and deserves thorough study. We do not aim to touch on it
in this work, but hope to consider it in future publications.

In parallel to our own work, Kofman, Linde and Starobinsky have also been
working on these issues for several years. Their results were recently
summarized in Ref. \REF\kls{L.Kofman, A.Linde and A.Starobinsky, UH-IfA-94/35,
SU-ITP-94-13, YITP/U-94-15, and HEP-TH-9405187, Submitted to: {\it Phys. Rev.
Lett.} (1994).} \kls. In Section 7 we will compare our results with those of
Ref. \kls.

\medskip
{\bf \chapter{Inflaton field dynamics}}

In this section we will derive approximate analytic expressions for the
evolution of the inflaton field as it undergoes quasiperiodic motion in the
expanding universe, and compute the resulting effective equation of state
of the inflaton stress-energy.

Consider the chaotic inflation scenario based on the scalar (inflaton) field
dynamics. The Lagrangian of the model is
$$
L = {1 \over 2} \, (\nabla \varphi)^2 \, - \, V(\varphi) \, - \,
{M_{\rm P}^2 \over 16\pi} \, R \, , \eqno  (1)
$$
with the inflaton field potential
$$
V(\varphi) = \lambda \mu^{4-q}\,|\varphi|^q \, , \eqno  (2)
$$
where $q$ is an arbitrary positive power. We are working in the standard
system of units in which $\hbar = c = 1$, and $M_{\rm P}$ denotes the Planck
mass.

The equations of motion for a homogeneous isotropic universe and for a
homogeneous scalar field look like
$$
H^2 + {\kappa \over a^2} = {8 \pi \over 3M_{\rm P}^2} \, \left({1 \over 2}
\, \dot \varphi^2 + V(\varphi)\right) \, , \eqno  (3)
$$
$$
\ddot \varphi + 3H \dot \varphi + V^\prime (\varphi) = 0 \, , \eqno  (4)
$$
where $H \equiv \dot a / a$ is the Hubble parameter, $\kappa = 0,\,\pm 1$
corresponds to different signs of spatial curvature, dots denote derivatives
with respect to the cosmological time $t$, and prime denotes derivative with
respect to $\varphi$.

  From the equation (4) it follows that
$$
\dot \varphi = a^{-3}\,\left({\rm const} - \int V^\prime (\varphi) a^3 dt
\right) \, . \eqno  (5)
$$
This equation describes both the rapid evolution of $\varphi$, for which
$\varphi
\propto a^{-3}$, and the regime of slow rolling-down during inflation, when
$\varphi \approx - (1/3H)\,V^\prime(\varphi)$.

According to the inflation scenario, during inflation the scalar field
$\varphi$ is rolling down the slope of its potential from its relatively large
value towards its minimum. The conditions for slow rolling and
for inflation are, respectively,
$$
|\ddot \varphi| \ll 3H|\dot \varphi| \, . \eqno   (6)
$$
and
$$
|\dot H| \ll H^2 \, , \eqno  (7)
$$

During inflation the term $\kappa / a^2$ in the left-hand side of Eq.(3) soon
becomes insignificant, and the conditions (6) and (7) read, respectively,
$$
\left(\sqrt {V(\varphi)}\right)^{\prime\prime} \ll {12 \pi \over M_{\rm P}^2}
\sqrt {V(\varphi)} \, , \eqno  (8)
$$
$$
\left(V^\prime (\varphi)\right)^2 \ll {24 \pi \over M_{\rm P}^2} V^2(\varphi)
\, . \eqno  (9)
$$
For the scalar field potential (2) these two conditions become respectively
$$
\varphi^2 \gg {q|q-2| \over 48 \pi} M_{\rm P}^2 \, , \eqno  (10)
$$
$$
\varphi^2 \gg {q^2 \over 24 \pi} M_{\rm P}^2 \, . \eqno  (11)
$$
Note that conditions (10) and (11) are essentially the same.

As the scalar field evolves towards its smaller values, the conditions (10)
and (11) cease to be valid, and a new regime for the scalar field
begins, namely, the regime of quasiperiodic evolution with decaying amplitude.
To describe this new cosmological period, it is convenient to rewrite the
system of equations (3), (4) (with $\kappa = 0$) as follows
$$
H^2 = {8 \pi \over 3M_{\rm P}^2} \rho \, , \eqno (12)
$$
$$
\dot \rho = - 3H \dot \varphi^2 \, , \eqno (13)
$$
where
$$
\rho \ = \ {1 \over 2} \, \dot \varphi^2 \, + \, V(\varphi) \, \eqno (14)
$$
is the scalar field energy density. Introducing the positive value
$\varphi_0(t)$ by the relation
$$
V(\varphi_0(t)) \ = \ \rho(t) \, , \eqno (15)
$$
we can present the evolution of the scalar field in the form
$$
\varphi(t) \ = \ \varphi_0(t) \cos \int W(t) dt \, , \eqno (16)
$$
where $W(t)$ is some unknown function of time. Note that the representation
(16)
for the scalar field evolution with $\varphi_0$ defined by (15) is always
possible. Using the equations (2) and (12)-(15), one can then derive the
following exact expression for the function $W(t)$
$$
W \ = \ \sqrt{{2(\rho - V(\varphi)) \over \varphi_0^2 - \varphi^2}} \,
\left(1\,\, \pm \,\, {6 H \varphi \over q \sqrt{2 \rho}}\,
\sqrt{1 - {V(\varphi) \over \rho}} \right) \, . \eqno (17)
$$
The signs ``$+$'' and ``$-$'' correspond respectively to the cases $\dot
\varphi
> 0$ and $\dot \varphi < 0$.
The second term in the brackets in (17) is much less than unity if
$$
\varphi_0^2 \ \ll \ {q\over 48 \pi}\left({q + 2 \over2}\right)^{{q + 2 \over
q}} \, M_{\rm P}^2 \, . \eqno (18)
$$
This condition is just the opposite of (10), (11). When (18) is valid, the
expression (17) for $W(t)$ acquires the simple approximate form
$$
W \ \approx \ \sqrt{{2(\rho - V(\varphi)) \over \varphi_0^2 - \varphi^2}}
\, , \eqno (19)
$$
and, in addition, the following condition becomes valid
$$
|\dot \varphi_0 / \varphi_0 | \ \ll \ W \, , \eqno (20)
$$
which allows us to regard the evolution of $\varphi(t)$ as quasiperiodic
with slowly decaying amplitude $\varphi_0(t)$. Henceforth we assume that the
condition (18) (hence also (19) and (20)) is satisfied.

Our aim now will be to derive approximate evolution equations for the values
$\rho(t)$ and $\varphi_0(t)$ under the condition (18). First we average the
equation of motion (13) over the time period $T$ of quasiperiodic motion
of $\varphi$. Rewriting (13) in accord with (14) as
$$
\dot \rho \ = \ - 6H \left(\rho - V(\varphi)\right) \, , \eqno (21)
$$
and taking the average of both sides we get
$$
{\Delta \rho(T) \over T} \ = \ - {6 \over T} \int \limits_0^T H\, \left(\rho -
V(\varphi)\right) dt \, , \eqno (22)
$$
where $\Delta \rho(T)$ is the change of $\rho$ over the time period $T$.
Now take into
account that according to (20) and to the equations (15) and (12) the values
$\varphi_0(t)$, $\rho(t)$ and $H(t)$ change only insignificantly during one
period of oscillation of the $\varphi$-field. This enables us to replace
all variables except $\varphi$ under the integral in (22) by their averaged
values. We will have then, to a good approximation
$$
{\Delta \rho(T) \over T} \ \approx \ - {6 \over T} \,  H \int \limits_0^T
\left( \rho - V(\varphi)\right) dt \, , \eqno (23)
$$
where $H$ and $\rho$ in (23) and below denote the corresponding averaged
values. The integral in (23) can be evaluated as
$$
{1 \over T} \int \limits_0^T \left(\rho - V(\varphi)\right) dt
\ \approx \
{\int \limits_{-\varphi_0}^{\varphi_0} \sqrt{\rho - V(\varphi)} d\varphi
\over \int \limits_{-\varphi_0}^{\varphi_0} {d\varphi \over \sqrt{\rho -
V(\varphi)}}} \  \approx
\  {{1 \over 2}\int \limits_{-\varphi_0}^{\varphi_0}
{\varphi V^\prime(\varphi) d\varphi \over \sqrt{\rho - V(\varphi)}}
\over \int \limits_{-\varphi_0}^{\varphi_0} {d\varphi \over \sqrt{\rho -
V(\varphi)}}} \  \approx
\  {q \over q + 2}\,  \rho \,  . \eqno (24)
$$
In (24) we took into account the relation (14), integrated in the
numerator by parts and made use of the equation (2).
In the approximation considered we can replace the finite difference
expression on the left-hand-side of (23) by the derivative. Equation (23)
then becomes
$$
\dot \rho \ \approx \ - {6q \over q + 2} \, H \, \rho \, . \eqno (25)
$$
All the values in (25) are now to be regarded as averaged in the sense
described above. Equation (25) is valid on time scales large
compared to the period of quasioscillations of $\varphi(t)$.
This implies the following effective equation of state for the
matter described by the scalar field $\varphi$
$$
p \, = \, {q - 2 \over q + 2}\, \rho \, . \eqno (26)
$$
In particular, for the most interesting cases $q = 2$ (quadratic potential)
and
$q = 4$ (quadric potential) we obtain, respectively, $p = 0$ (dust) and
$p = \rho / 3$ (radiation).

Using the definition (15) we immediately obtain the approximate equation
for the evolution of the value $\varphi_0(t)$
$$
\dot \varphi_0 \ \approx \ - {6 \over q + 2} \, H\, \varphi_0 \, . \eqno (27)
$$
\medskip
{\bf \chapter{Particle production: Born approximation}}

In this section we will study the process of particle production by the
rapidly oscillating inflaton field. We take the interaction Lagrangian
to be
$$
L_{\rm int} \ = \ - f\varphi \overline \psi \psi - \left(\sigma \varphi +
h \varphi^2 \right) \chi^2 \, , \eqno (28)
$$
where $\overline \psi$ and $\psi$ describe spinor particles and $\chi$
describes scalar particles with corresponding masses $m_\psi$
and $m_\chi$, $f$ and $h$ are dimensionless coupling constants,
and $\sigma$ is a coupling constant of dimension of mass.
We will treat the scalar field $\varphi$ as a classical
external field. In this section we will employ ordinary perturbation theory
in
coupling constants, that is, we will work in the Born approximation.
To calculate the rate of particle production we first develop
the quasiperiodic evolution of the scalar field $\varphi$ into harmonics
$$
\varphi(t) \ = \ \sum_{n=1}^\infty \varphi_n  \cos (n \omega t) \, , \eqno (29)
$$
$$
\varphi^2(t) \ = \ \overline{\varphi^2} + \sum_{n=1}^\infty \zeta_n
 \cos (2n \omega t) \, , \eqno (30)
$$
where the value $\overline{\varphi^2} \
\approx \ \varphi_0^2/2$ which is slowly varying with time is $\varphi^2$
averaged over the rapid
oscillations of the scalar field $\varphi$. $\varphi_n$, $\zeta_n$ are
amplitudes which are slowly varying with time, and
$$
\omega \, = \, {2\pi \over T} \, = \, c \, {\sqrt \rho \over \varphi_0}
\, \eqno (31)
$$
is the leading frequency, also slowly varying with time, related to the
period $T$ of the oscillations of the field $\varphi$. The constant $c$ in
(31) is of order 1, and is given by
$$
c \, = \, \pi \sqrt 2 \, \Big/ \int\limits^1_{-1} {dx \over \sqrt{1 - x^q}}
\, \eqno (32)
$$
in the case of a scalar field potential given by (2). To derive the estimate
(31) we made use of the relation (14).

Assuming the condition (18), it follows that the oscillation period of
$\varphi$ is small compared to the Hubble time,
$$
H \, \ll \,  \omega \, \eqno (33)
$$
This last condition allows us to disregard the effect of cosmological
expansion in evaluating particle production rates. We will also assume that
particle masses and the coupling constant $h$ are sufficiently small, so that
$$
m_\psi^2\, ,\,\,\, m_\chi^2 \, + \,
h\overline{\varphi^2} \ \ \ll \ \ \omega^2  \, . \eqno (34)
$$

The rates of particle production, that is, the total number of pairs produced
per unit
volume and unit time are then given in first order perturbation theory in
coupling constants $f$, $\sigma$ and $h$ by the following relations
$$
w_{\overline\psi\psi} \, = \, {f^2 \over 16\pi} \sum_{n=1}^\infty
\left(\varphi_n n \omega \right)^2 \, , \eqno (35)
$$
$$
w_{2\chi} \, = \, {1 \over 16\pi} \left( \sigma^2\sum_{n=1}^\infty \varphi_n^2
\, + \, h^2\sum_{n=1}^\infty \zeta_n^2 \right) \, , \eqno (36)
$$
Remembering the expansions (29) and (30) we can put the sums in the last
expressions into the following form
$$
\sum_{n=1}^\infty \left(\varphi_n n \omega \right)^2 \, = \,
2<\dot \varphi^2> \, , \eqno (37)
$$
$$
\sum_{n=1}^\infty \varphi_n^2 \, = \, 2<\varphi^2> \, , \eqno (38)
$$
$$
\sum_{n=1}^\infty \zeta_n^2 \, = \, 2<\left(\varphi^2 - \overline{\varphi^2}
\right)^2> \, , \eqno (39)
$$
where the brackets $<\ldots>$ denote the average over
quasioscillations of the scalar field $\varphi(t)$. These average values
in the expressions (37)-(39) can be estimated in exactly the same manner in
which the
equation (24) is derived. We obtain, finally, the following relations for
the particle production rates
$$
w_{\overline\psi\psi} \, = \, {f^2 \over 4\pi}\,{q \over q + 2}\, \rho
\, , \eqno (40)
$$
$$
w_{2\chi} \, = \, {1 \over 8\pi} \left( c_\sigma\, \sigma^2  \varphi_0^2 \,
+ c_h \, h^2  \varphi_0^4 \right)\, , \eqno (41)
$$
where
$$
c_\sigma \, = \, \int\limits_{-1}^1 {x^2dx \over \sqrt{1 - x^q}} \, \Big/
\int \limits_{-1}^1 {dx \over \sqrt{1 - x^q}} \, , \eqno (42)
$$
$$
c_h \, = \, \int\limits_{-1}^1 {\left(x^2 - c_\sigma^2\right)^2dx \over
\sqrt{1 - x^q}}
\Big/ \int \limits_{-1}^1 {dx \over \sqrt{1 - x^q}} \, , \eqno (43)
$$
are constants of order unity.

Higher order contributions to the particle production rates (40)-(41) will
be negligible if
$$
f^2\, , \,\,\, h \, , \,\,\, {\sigma \over \varphi_0} \,\, \ll \,\,
\left({\omega \over \varphi_0} \right)^2 \, . \eqno (44)
$$
Hence, our assumption (34) for the smallness of the coupling constant $h$ is
justified since $\overline{\varphi^2}\ \approx \ \varphi_0^2/2$.
We will see below what the conditions (44) imply for the concrete inflaton
potentials.

Equations (40)-(41) enable us to estimate the energy density production in
the form of $\overline\psi$, $\psi$ and $\chi$ particles created.
To do this, note that pairs of particles produced have average energy of the
order $\omega$, if we reasonably assume that the sums on the right-hand sides
of (35), (36) rapidly converge. Then the equation for $\rho_p$,
the particle energy density produced, can be written as follows
$$
\dot \rho_p \, ({\rm production})\, = \, \left(\Gamma_\psi +
\Gamma_\chi \right)  \rho \, , \eqno (45)
$$
where
$$
\Gamma_\psi \ = \ {\cal O}(1)\,{f^2 \over 4\pi}\, \omega \, , \eqno (46)
$$
is the contribution corresponding to the production of fermionic particles,
and
$$
\Gamma_\chi \ = \ {\cal O}(1)\, {\sigma^2
\over 4\pi \omega} \, + \, {\cal O}(1)\, {(h \varphi_0)^2 \over 4\pi \omega}
\, , \eqno (47)
$$
is the contribution corresponding to the production of scalar particles
$\chi$. Here and below we denote by the symbol ${\cal O}(1)$ a constant
of order
unity. The values $\Gamma_\psi$ and $\Gamma_\chi$ can be interpreted as the
``$\varphi$-particle'' decay rates in corresponding decay channels.
\medskip
{\bf \chapter{Particle production: parametric resonance}}

The expressions (45)-(47) for the particle energy density production rates
have been obtained
in first order Born approximation, and in this form they have been used
in all the
seminal papers on reheating. However, as it has been shown in [\tb], this
approximation does not take appropriate account of the parametric resonance
phenomena
which take place during the oscillation period of the scalar field $\varphi$.
Parametric resonance effects can be shown to be insignificant for fermionic
fields [\dk]. The estimate (46) for fermion production made in the preceding
section thus remains valid.
However the question of the parametric resonance effect for the scalar
field $\chi$
has to be considered and this will be the aim of this chapter. We will show
that the resonance occurs when the frequency $\omega_k$ of the quantum field
mode is equal to the half-integer multiples of the inflaton frequency,
$\omega_k^2 \approx \left({n \over 2} \omega \right)^2$. This results in
exponentially enhanced production of particles in narrow resonance bands, at
rates which are computed below.

The evolution of a particular mode $\chi_k$ of the quantum scalar field
$\chi$
in the presence of the quickly oscillating classical scalar field $\varphi$
with the interaction (28) is described by the following equation
$$
\ddot\chi_k + 3H\dot\chi_k + \left(\underline k^2 + m_\chi^2 + 2\sigma \varphi
+ 2h\varphi^2\right)\chi_k = 0 \, , \eqno (48)
$$
where $\underline k = k/ a$ is the physical wavenumber of the mode under
consideration, $k$ is the comoving number.

Performing the transformation
$$
\chi_k = {Y_k \over a^{3/2}} \eqno (49)
$$
one obtains from the equation (48) the following equation for the function
$Y_k$:
$$
\ddot Y_k + \left(\omega_k^2(t) + g(\omega t)\right)Y_k = 0
\, , \eqno (50)
$$
where
$$
\omega_k^2(t) \, = \, \underline k^2 + m_\chi^2 - {9 \over 4} H^2 -
{3 \over 2} \dot H + 2h\overline{\varphi^2} \, , \eqno (51)
$$
and
$$
g(\omega t) \, = \, 2\sigma \varphi + 2h\left(\varphi^2 - \overline
{\varphi^2} \right) \, . \eqno (52)
$$
The function $g(\omega t)$ is to a good approximation a
$2\pi/\omega$-periodic function of time $t$ with two leading frequencies
$\omega$ and $2\omega$, as can be
seen from (52) and the expansions (29), (30).
We will regard this function as a small perturbation in the equation (50),
and to be able to do this we require the condition
$$
2\sigma\varphi_0 + h\varphi_0^2 \, \ll \, \omega_k^2 \, . \eqno (53)
$$
which we will assume to be valid.

The time dependence of the coefficients in the equations (48), (50) will lead
to
$\chi$-particle production which can be described by the standard Bogolyubov
transformation technique. The problem consists in finding the solution to the
equation (50) with initial conditions corresponding to the initial
vacuum state and then evaluating the Bogolyubov transformation coefficients.
We expect that for the equation (50) parametric resonance effect will be
dominant hence we are going to study this effect in detail.

  From the general theory of parametric resonance (see Appendix A for details
\footnote{1)}{In Appendix A we
explicitly introduced a small dimensionless parameter $\epsilon$. In the main
text we have absorbed $\epsilon$ into the definition of $g$.})
one knows
that parametric resonance occurs for the equation (50) for
certain values of the frequencies $\omega_k$. Namely, the resonance
in the lowest frequency resonance band occurs for those values of
$\omega_k$ for which
$$
\omega_k^2 - \left({n \over 2}\, \omega \right)^2 \, \equiv \
\Delta_n \ < \ |g_n| , \eqno (54)
$$
where $n$ is an integer, and $g_n$ is the amplitude of the $n$-th Fourier
harmonic of the function $g(\omega t)$. Of course, for such a conclusion to
be valid in our case when the frequencies $\omega_k$ and $\omega$
depend on time we must be sure that these frequencies
change with time slowly as compared to their own values, that is
$$
\Big|{\dot \omega_k \over \omega_k}\Big| \, , \,\,\, \Big|{\dot \omega
\over \omega} \Big|
\,\, \ll \,\, \omega_k \, ,  \,\,\, \omega \, . \eqno (55)
$$
But this condition immediately follows from the estimates (20), (33),
and from (53) if we also take the relation (54) into account. Note that
condition (53) for $\omega_k$ in the resonance band coincides
with the conditions (44) on the validity of Born approximation.

In this paper we will consider only resonance in the lowest resonance
frequency band for each Fourier harmonic of the function $g$ as
given by equation (54). The
effects of higher resonance bands, as is well known, are of higher order
in the amplitudes $|g_n|$ which we assumed to be small. Hence we expect these
effects to be less efficient.

  From the conditions (20), (33), (34) and
(53) it also follows that part of the expression (51) satisfies
$$
\omega_k^2 - \underline k^2 \ \equiv \  m_\chi^2 - {9 \over 4} H^2 -
{3 \over 2} \dot H + 2h\overline{\varphi^2} \ \ll \  \omega^2 \, . \eqno (56)
$$
It then follows from the equation (54) that
$$
\underline k^2 \, \gg \, m_\chi^2 - {9 \over 4} H^2 - {3 \over 2} \dot H +
2h\overline{\varphi^2} \, \eqno (57)
$$
for the resonance values of $\underline k^2$. This means that the redshift
of the frequency $\omega_k$ due to the expansion of the universe is, to a good
approximation,
$$
\omega_k(t) \, \propto \, \underline k \, \propto \, a^{- 1}(t) \, . \eqno (58)
$$

Due to the condition (33) we can neglect this cosmological redshift on the
timescale of the scalar field $\varphi$ oscllations. The mean occupation
numbers for the generated $\chi$-particles are then given by the Bogolyubov
coefficients $\beta_k$ as $N_k = |\beta_k|^2$ and the expression for them can
be approximately written as follows (see Appendix B for derivation\footnote{1)}
{We must use the expression (B32) for the number of particles produced,
because in our case the value $\mu_+$ varies with time.})
$$
N_k  \, \simeq  \sinh^2 \left(\int \mu_+\, dt \right)
\, , \eqno (59)
$$
where
$$
\mu_+ \, = \, {1 \over n\omega} \sqrt{|g_n|^2 - \Delta_n^2}
\, \eqno (60)
$$
is the eigenvalue of the growing resonance mode, and $\Delta_n$ is given by
(54).

The approximation (59) is valid only if we can treat the value $\mu_+$
as adiabatic. For our purposes it is sufficient to consider values of
$\mu_+$ close to the resonance (when $\Delta_n = 0$) and the
variation of $\mu_+$ with time is mostly due to the expansion of the
universe.
Then the adiabaticity condition (see the equation (B31)) will be as follows
$$
\mu_+ \ \gg \ H \, . \eqno (61)
$$

As time goes on, the mode with the particular wavenumber $k$ gets out of
the resonance band, and the value $N_k$ ceases to grow. The reason for
the frequency $\omega_k$ (initially in resonance) to get out of the resonance
band is that the redshift evolution (58) of the frequency $\omega_k$ is
different from the evolution of the leading frequency $\omega$ and hence
the resonance frequency $n \omega /2$ is continuously passing through
the spectrum of frequencies $\omega_k$ of the modes of the scalar field
$\chi$. (In the case of $\lambda \varphi^4$ potential, the value $\omega$
does redshift at the same rate, and one has to take into account the
back-reaction of the created particles to determine this frequency shift;
see Section 5.)

Denote by $\omega_{\rm res}$ the resonance frequency
$n\omega / 2$ and by $\Delta\omega_k$ denote the difference
$\omega_k - \omega_{\rm res}$. Both frequencies $\omega_{\rm res}$ and
$\omega_k$ change with time. In a small time $\delta t$ the shift between
these frequencies will be $\delta \omega =
|d\Delta \omega_k / d t|_{\Delta \omega_k = 0} \delta t$. Hence a new region
in phase space of volume $\omega_{\rm res}^2 \delta \omega / 2 \pi^2$
will be filled with particles. The average number of particles in every state
in this phase space region will be given by $N_k$ and each particle will have
energy $\omega_{\rm res}$. Then taking sum over all parametric
resonance bands (with various numbers $n$ in (54)) we can write
down the equation for the energy density $\rho_\chi$ of the scalar particles
produced in the following form
$$
\dot \rho_\chi \, {\rm (resonance \,\, production)}\,\, = \,
\sum_{\rm all\ resonance\ bands} \,
{1 \over 2\pi^2} \, N_{\rm res} \, \omega_{\rm res}^3 \, \Big|
{d\Delta\omega_k \over dt}\Big|_{\Delta\omega_k = 0} \, , \eqno (62)
$$
where $N_{\rm res}$ is the maximal value of $N_k$, for the current
resonance value of $k$, achieved after the
corresponding mode has passed through the resonance band and has been
amplified.

Under condition (34) particles produced will be ultrarelativistic and
hence their contribution to energy density will decrease due to the
cosmological
expansion. Taking also this process into account we are able to write down
the following complete equation for the evolution of the energy density
$\rho_p$ of
the particles produced (including the contribution from fermions)
$$
\dot \rho_p \, = \, -\,4H \rho_p \, + \, \left(\Gamma_\psi
+ \Gamma_\chi^{\rm (res)} \right) \rho \, , \eqno (63)
$$
where
$$
\Gamma_\chi^{\rm (res)} \, = \, {1 \over 2\pi^2\rho} \, \sum_{\rm all\
resonance\ bands} \,
N_{\rm res} \, \omega_{\rm res}^3 \, \Big|
{d\Delta\omega_k \over dt}\Big|_{\Delta\omega_k = 0} \,  \eqno (64)
$$
plays the role analogous to that of $\Gamma_\chi$ in (45). The values
of $\Gamma_\chi^{\rm (res)}$
given by (64) and $\Gamma_\chi$ given by (47) are to be
compared in order to conclude about the significance of the parametric
resonance for the process of scalar particle production.
In what
follows we consider two important cases of the inflaton potential
$V(\varphi)$, namely, quartic and quadratic potentials.
\medskip
{\bf \chapter{Case of potential $ V(\varphi) = \lambda\varphi^4$}}

In this case we have from (31)
$$
\omega \, = \, c \, \sqrt \lambda \varphi_0 \, , \eqno (65)
$$
where the constant $c$ is given by (32) with $q = 4$.
The resonance frequency $\omega_{\rm res}$ is thus proportional to
$\varphi_0$ and
by virtue of (27) in the absence of back reaction of the particle
production process on the scalar field $\varphi$ we would have
$$
\omega_{\rm res} \, \propto \, a^{-1} \, . \eqno (66)
$$
As we also have the condition (58) this would mean that the resonance
amplification takes place for all times in the same wavebands, which satisfy
the condition (54). However due to the energy loss due to particle production,
the scalar field
amplitude $\varphi_0$, and hence the frequency $\omega_{\rm res}$, will
decrease more
rapidly than given by (66). Wee see that in order to make correct
estimates of particle production in the model with quartic inflaton
potential one has to take into account back-reaction on the inflaton
evolution.

Let us then consider back-reaction of the particle production
process on the evolution of the scalar field $\varphi$. Since particle
production is a random quantum mechanical phenomenon, the change in the
evolution of the scalar field $\varphi$ will be random. We can picture
the back-reaction as leading to abrupt changes in the state of
the scalar field at those random moments of time at which particles are
produced. On sufficiently
large time scales, on which many particles are produced, this change will
exhibit regular features, and it is these features which we are to determine.

The evolution of the scalar field $\varphi$ can be described by two
parameters,
namely, the scalar field energy density $\rho$ and the frequency $\omega$
of its
quasioscillations. In absence of back-reaction both these parameters are
changing slowly with time, when compared to the time scale $T = 2\pi/\omega$
of the quasioscillations. The value $\omega$ is directly related to the
energy density through the relation (31). We find it possible to describe
the state of the
scalar field in terms of these two parameters also if we take the
back-reaction mentioned above into account. Indeed, it
is always possible to speak of the scalar field energy density. As far
as its quasioscillation frequency is concerned we note that its very
existence is necessary for particle production, with rates given by
(40)-(41), or with average numbers (59) in the case of parametric resonance.

Also note that on the
quasioscillation time scale there can be but few particles produced (this
fairly general statement follows from the standard analysis of transition
amplitudes). Hence random moments of the abrupt changes in the scalar field
state will be time-separated in average by more than
one scalar field quasioscillation period. Between these moments of abrupt
changes the scalar
field is evolving unperturbed. From this it follows
that we can assume the validity of the relation (31) between the
quasioscillation frequency
$\omega$ and the energy density $\rho$.

To determine in the most simple way
the changes in the evolution of the scalar field energy density $\rho$ we
can apply the energy conservation law. The energy is transferred from
the scalar field $\varphi$ to the particles produced. Hence, given the energy
production
rate (63) of the latter we thereby know the energy loss of the scalar field
$\varphi$. The equation for the inflaton energy density will then be
$$
\dot \rho \, = \, - \, (4 H + \Gamma_\psi + \Gamma_\chi^{\rm (res)}) \, \rho \,
, \eqno (67)
$$
with $\Gamma_\chi^{\rm (res)}$ given by (64). From the equation (2) with $q =
4$ and from (15)
it then follows
$$
\dot \varphi_0 \, = \, - \, \left(H + {1 \over 4} \left(\Gamma_\psi +
\Gamma_\chi^{\rm (res)}
\right) \right) \,\varphi_0 \, . \eqno (68)
$$
Now taking (58) and (65) into account we will obtain the following expression
for the time derivative factor in (62)
$$
\Big| {d\Delta\omega_k \over dt} \Big|_{\Delta\omega_k = 0} \, =
\, {1 \over 4} \left(\Gamma_\psi + \Gamma_\chi^{\rm (res)} \right)\,
\omega_{\rm res} \, , \eqno (69)
$$
and from the equation (64) we then obtain
$$
\sum_{\rm all\ resonance\ bands} \, N_{\rm res}\, \omega_{\rm res}^4 \, = \,
8 \pi^2 \rho \left(1 + {\Gamma_\psi \over \Gamma_\chi^{\rm (res)}} \right)^{-
1} \, . \eqno (70)
$$

It is very remarkable that in spite of all the details of the interaction
between
the fields $\varphi$ and $\chi$ the maximal occupation
number mean values depend through (70) only on the resonance
frequencies
$\omega_{\rm res}$ in different resonance bands. This significant feature
of the quartic inflaton
potential makes the total analysis rather simple.

It remains to estimate the value $\Gamma_\chi^{\rm (res)}$. The expression for
$N_k$ is given by (59) where the integration in the argument of the hyperbolic
sine is taken over the time during which
resonance is taking place for the mode with the wavenumber $k$.
In terms of the value $\Delta\omega_k
= \omega_k - \omega_{\rm res}$, the expression (60) for $\mu_+$ can be
rewritten as follows
$$
\mu_+ \, = \, {1 \over 2 \omega_{\rm res}} \sqrt{|g_n|^2  -
\Delta\omega_k^2(\Delta\omega_k + 2\omega_{\rm res})^2} \, . \eqno (71)
$$
Then using (69) and taking into account the smallness of the value
$|g_n|$ as compared to $\omega_{\rm res}^2$ (this follows from the relations
(52) and (53)) we can estimate the integral in the exponent of (59) as follows
$$\eqalign{
\int \mu_+ dt \, &\approx \,  \int\limits_{- |g_n|/2\omega_{\rm
res}}^{|g_n|/2\omega_{\rm res}}
{1 \over 2 \omega_{\rm res}} \sqrt{|g_n|^2  -
\Delta\omega_k^2(2\omega_{\rm res})^2}\,\, {4 \over \left(\Gamma_\psi +
\Gamma_\chi^{\rm (res)}\right)\,
\omega_{\rm res}}\, d \Delta\omega_k \,  \approx \cr
& \approx \,\,
{|g_n|^2 \over \left(\Gamma_\psi + \Gamma_\chi^{\rm (res)}\right)\,
\omega_{\rm res}^3}
\, \int \limits_{-1}^1 \sqrt{1 - x^2}\, dx \, = \, {\pi \over 2} \,
{|g_n|^2 \over \left(\Gamma_\psi + \Gamma_\chi^{\rm (res)}\right)\,
\omega_{\rm res}^3}  \, . \cr} \eqno (72)
$$
In deriving (72) we considered slowly changing variables as constants.
Then the expression (59) for $N_{\rm res}$ becomes
$$
N_{\rm res} \, \simeq  \sinh^2\left( {\pi |g_n|^2 \over 2\left(\Gamma_\psi +
\Gamma_\chi^{\rm (res)}\right)\,
\omega_{\rm res}^3} \right) \, . \eqno (73)
$$
Substituting this expression for $N_{\rm res}$ into (70) we obtain the
following implicit expression for the value $\Gamma_\chi^{\rm (res)}$
$$
\sum_{\rm all\ resonance\ bands} \omega_{\rm res}^4  \sinh^2\left( {\pi
|g_n|^2 \over 2 \left(\Gamma_\psi + \Gamma_\chi^{\rm (res)}\right)\,
\omega_{\rm res}^3}
\right)\,  \simeq \, 8\pi^2\rho
\left(1 + {\Gamma_\psi \over \Gamma_\chi^{\rm (res)}} \right)^{- 1}
\, . \eqno (74)
$$

Consider first the case when energy production of scalar particles dominates
over that of fermions so that
$$
\Gamma_\psi \ < \ \Gamma_\chi^{\rm (res)} \, . \eqno (75)
$$
Typically one of the terms, say, of the $n$-th resonance band dominates
in the sum in the left-hand side of (74). In the case (75) $\Gamma_\chi^{\rm
(res)}$ is determined
by this dominant term as follows
$$
\Gamma_\chi^{\rm (res)} \, \simeq \, {\pi |g_n|^2 \over \omega_{\rm res}^3} \,
\log^{- 1} \left(
{32\pi^2\rho \over \omega_{\rm res}^4} \right) \, . \eqno (76)
$$

Two of the Fourier harmonics of the function $g$ given by (52) will be
of appreciable value, namely,
$$
|g_1| \, \simeq \, \sigma\varphi_0 \, , \hskip2cm \left({\rm with} \ \
\omega_{\rm res} \, = \, {\omega \over 2}\right) \, , \eqno (77)
$$
and
$$
|g_2| \, \simeq \, {h\varphi_0^2 \over 2} \, , \hskip2cm \left({\rm with} \ \
\omega_{\rm res} \, = \, \omega \right) \, . \eqno (78)
$$

If the contribution from $n = 1$ is dominant in (74) (that is if $\varphi_0 <
\sigma/h$) then from (76), (77) and using the expression (65) for
$\omega$ and the explicit expression for the inflaton energy density
$\rho = \lambda \varphi_0^4$ we get
$$
\Gamma_\chi^{\rm (res)} \, \simeq \, {8 \, \pi \, \over c^3 \, \lambda^{3/2}}
\,\, {\sigma^2 \over
\varphi_0}\
\log^{- 1} \left( {512 \, \pi^2 \over c^4 \, \lambda}\right) \, ,
\hskip2cm {\rm for} \ \varphi_0 < {\sigma \over h} \, . \eqno (79)
$$
In this case the adiabaticity condition (61) reads
$$
\varphi_0^2 \ \ll \ {M_{\rm P}\, \sigma \over \lambda} \, . \eqno (80)
$$

If the contribution from $n = 2$ is dominant in (74) then from (76), (78)
we in a similar way obtain
$$
\Gamma_\chi^{\rm (res)} \, \simeq \, {\pi \, h^2 \over 4 c^3 \, \lambda^{3/2}}
\,\, \varphi_0  \
\log^{- 1} \left( {32 \, \pi^2 \over c^4 \, \lambda}\right) \, ,
\hskip2cm {\rm for} \ \varphi_0 > {\sigma \over h} \, . \eqno (81)
$$
In this case the adiabaticity condition (61) implies
$$
\varphi_0 \ \ll \ M_{\rm P}\, {h \over \lambda} \, . \eqno (82)
$$

According to the note made at the end of the previous section, the values (79)
and (81)
for $\Gamma_\chi^{\rm (res)}$ are to be compared to the values of $\Gamma_\chi$
in the corresponding
cases. The expression for $\Gamma_\chi$ is given by (47). Calculating the
ratios of $\Gamma_\chi^{\rm (res)}$ to $\Gamma_\chi$ in the two cases discussed
just above we obtain
$$
{\Gamma_\chi^{\rm (res)} \over \Gamma_\chi} \  \simeq \ {32 \, \pi^2 \over c^2
\, \lambda} \ \log^{-1}
\left({512 \, \pi^2 \over c^4 \, \lambda}\right) \, , \hskip2cm
{\rm for} \ \varphi_0 < {\sigma \over h} \, , \eqno (83)
$$
$$
{\Gamma_\chi^{\rm (res)} \over \Gamma_\chi} \  \simeq \ {\pi^2 \over c^2 \,
\lambda} \ \log^{-1}
\left({32 \, \pi^2 \over c^4 \, \lambda}\right) \, , \hskip2cm
{\rm for} \ \varphi_0 > {\sigma \over h} \, . \eqno (84)
$$

We see that these ratios look similar in two cases considered, and for typical
values of the inflaton coupling constant [\l]
$$
\lambda \ \sim \ 10^{- 12} \, \eqno (85)
$$
they are extremely large. Hence, for the inflaton field with the potential
$\lambda \varphi^4$ with small $\lambda$, in the domain of applicability of
perturbation theory (given by (44)), parametric resonance effects strongly
dominate over the usual Born approximation estimates, so that the latter
represent a serious underestimation of the particle production rate.

Let us proceed further and estimate the possible reheating temperature for
the model
considered.\footnote{1)}{We take the reheating temperature to be the
temperature which
would correspond to the energy density of the particles produced. This
assumes that the particles produced will be thermalized soon after their
creation. In this paper we do not touch on the theory of the thermalization.}
The equation for the
energy density of the particles produced is (63). We should take into account
also the equation (67), modify the equation (12) as follows
$$
H^2 \ = \ {8 \pi \over 3 M_{\rm P}^2}\, \left(\rho + \rho_p
\right) \, , \eqno (86)
$$
and then solve the
system of equations (63), (67) and (86) with initial condition $\rho_p
= 0$ at the moment of the end of inflation. Then the maximal value
achieved in the course of evolution by $\rho_p$ would determine the
reheating temperature. The system of equations mentioned can be solved
exactly if we use the condition (75) to neglect $\Gamma_\psi$ and the
expressions (79) or (81) for $\Gamma_\chi^{\rm (res)}$. In the case (79)
the solution will be
$$
\rho \ = \ \rho_0 \left(1 - {\Gamma_0 \over 12 H_0} \left(
e^{3 \tau} - 1 \right) \right)^4 e^{- 4 \tau} \, , \eqno (87)
$$
and in the case (81) the solution is
$$
\rho \ = \ \rho_0 \left(1 + {\Gamma_0 \over 4 H_0} \left(
e^\tau - 1 \right) \right)^{- 4} e^{- 4 \tau} \, , \eqno (88)
$$
where $\Gamma_0$, and $H_0$ are the initial values of
$\Gamma_\chi^{\rm (res)}$ and $H$ respectively, $\rho_0$ and $a_0$ are the
initial values of the total energy density and the scale factor, and
$\tau = \log
\left(a / a_0\right)$. In both cases (87) and (88) the solution for $\rho_p$ is
given by
$$
\rho_p \ = \ \rho_0 e^{- 4 \tau} - \rho \, . \eqno (89)
$$

Until the condition $\rho_p \lsim \rho$ holds, the solutions
(87) and (88) can be approximately written as
$$
\rho \ \approx \ \rho_0 \left(1 - {1 \over 12} \left({\Gamma_\chi^{\rm (res)}
\over H} - {\Gamma_0 \over H_0} \right) \right)^4 e^{- 4 \tau} \, , \eqno (90)
$$
and
$$
\rho \ \approx \ \rho_0 \left(1 + {1 \over 4} \left({\Gamma_\chi^{\rm (res)}
\over H} - {\Gamma_0 \over H_0} \right) \right)^{- 4} e^{- 4 \tau}
\, , \eqno (91)
$$
respectively.

The solutions obtained determine the maximal value achieved by $\rho_p$.
Using the expressions (89), (90) and (91) it is easy to see that this
value is achieved as soon as the condition
$$
\Gamma_\chi^{\rm (res)} \ \gsim \ 4\,H \,  \eqno (92)
$$
becomes valid. After this particle creation proceeds very effectively and
the inflaton field loses all its energy in less then one Hubble time. We shall
therefore use the condition (92) to estimate the reheating temperature in
various cases. First, assume that reheating takes place in the region
$\varphi_0 > \sigma / h$. Then using the expression (81) for
$\Gamma_\chi^{\rm (res)}$ we obtain from (92)
$$
\varphi_0 \ \lsim \ 10^{- 1}\, M_{\rm P} \left({h \over \lambda}\right)^2\,
\log^{- 1} \, {1 \over \lambda} \, , \eqno (93)
$$
where we also used that $\lambda$ is small to simplify the logarithmic
factor. Now, the condition $\varphi_0 > \sigma/h$ assumed above will read
$$
\sigma \ \lsim \ 10^{- 1}\, M_{\rm P}\, h \left({h \over
\lambda}\right)^2\, \log^{- 1} \, {1 \over \lambda} \, . \eqno (94)
$$
Hence if the condition (94) is valid, the value of the inflaton field amplitude
$\varphi_0$ at the moment of reheating is given by the right-hand side of (93).
Note that the
adiabaticity condition (82) is valid for the values of $\varphi_0$ given
by (93) since
the ratio $h / \lambda$ has to be small according to the conditions (44).

If the condition (94) is not valid then reheating takes place in the region
$\varphi_0 < \sigma/h$ and we must use the expression (79) for
$\Gamma_\chi^{\rm (res)}$. In this case from (92) we obtain the estimate
$$
\varphi_0 \ \lsim \ {M_{\rm P}^{1/3}\, \sigma^{2/3} \over \lambda^{2/3}} \
\log^{- 1/3}\, {1 \over \lambda} \, , \eqno (95)
$$
and the condition $\varphi_0 < \sigma/h$ would imply, as it should, the
opposite of (94), namely
$$
\sigma \ \gsim 10^{- 1} \ M_{\rm P}\, h \left({h \over \lambda}\right)^2\,
\log^{- 1} \, {1 \over \lambda} \, . \eqno (96)
$$
In this case the last inequality together with (95) leads to the estimate
$$
\varphi_0 \ \gsim \ 10^{- 1}\, M_{\rm P} \left({h \over \lambda}\right)^2\,
\log^{- 1} \,
{1 \over \lambda} \,  \eqno (97)
$$
for the boundary value of $\varphi_0$ at which effective reheating starts.

Due to the conditions (44) the adiabaticity condition (80) can again be seen
to be valid for the values (95) of $\varphi_0$.

We see that the expression in the right-hand side of (93) gives the minimum
possible value for the inflaton
field amplitude $\varphi_0$ at the moment of reheating. We stress again that
for the
values of $\varphi_0$ obeying (93) the condition $4 H \lsim
\Gamma_\chi^{\rm (res)}$
holds and reheating is extremely effective. To estimate the
numerical value of the threshold (93) note, that according to the conditions
(44) (which we
remember to be necessary in order that perturbation theory is valid) we have
the condition $h \ll \lambda$. So taking for example
$$
{h \over \lambda} \ \simeq \ 10^{- 1} \, \eqno (98)
$$
and taking into account (85) we will get from (93)
$$
\varphi_0 \ \lsim \ 10^{- 4} M_{\rm P} \, . \eqno (99)
$$
The reheating temperature $T_{\rm rh}$ is then estimated from the condition
that the value of $\rho_p$ at the moment of reheating is of the order of the
value of $\rho$ at the moment when the condition (92) is first achieved,
and from the relation
$$
T_{\rm rh} \ \simeq \ \rho_p^{1/4} \, . \eqno (100)
$$
Finally we obtain
$$
T_{\rm rh} \ \sim \ \lambda^{1/4}\, \varphi_0 \ \sim \ 10^{- 1}\, M_{\rm P}\,
\lambda^{1/4}\left({h \over \lambda}\right)^2\, \log^{- 1} \,
{1 \over \lambda} \, . \eqno (101)
$$
In case of the values (85), (98) for the couplings we get
$$
T_{\rm rh} \ \sim \ 10^{- 7} M_{\rm P} \ \simeq \ 10^{12}\,
{\rm GeV} \, . \eqno (102)
$$

Now let us see what our condition (75) means. In the region $\varphi_0
> \sigma/h$ using the expressions (46) and (81) we obtain the condition
for the coupling $f$ (again simplifying logarithmic factor by using smallness
of the coupling constant $\lambda$)
$$
f^2 \ < \ {\cal O}(10)\, \left({h \over \lambda}\right)^2\, \log^{- 1}\,
{1 \over \lambda} \, , \eqno (103)
$$
which does not depend on the value of $\varphi_0$. For the values (98) of
the ratio $h / \lambda$ the condition (103) is not very restrictive.

In the region $\varphi_0 < \sigma / h$ using expression (79) we obtain
from (75) the estimate
$$
f^2 \ < \ {\cal O}(10^2)\, {1 \over \lambda^2}\, \left({\sigma \over \varphi_0}
\right)^2\,
\log^{- 1}\, {1 \over \lambda} \, , \eqno (104)
$$
which depends on the value of $\varphi_0$ but certainly is satisfied for
$\varphi_0$ sufficiently small. For the condition (104) to be valid for
our estimate (95) of the inflaton field amplitude at the moment of reheating
the coupling $f$ is to be
$$
f^2 \ < \ {\cal O}(10^2)\, \left({\sigma \over \lambda\, M_{\rm
P}}\right)^{2/3}\,
\log^{- 1/3}\, {1 \over \lambda} \, . \eqno (105)
$$

Thus, to conclude this part, if the value $\sigma$ satisfies (94) {\it and}
the coupling $f$ satisfies (103) then the value of the inflaton field
amplitude $\varphi_0$ at the moment of reheating is given by the right-hand
side of (93). If the
value $\sigma$ satisfies (96) {\it and} the coupling $f$ satisfies (105) then
the value of the inflaton field amplitude $\varphi_0$ at the moment of
reheating is given by the right-hand side of (95) and satisfies (97). In both
cases the reheating temperature
is not less then given by (101), with the numerical estimates (102) for the
coupling values (85) and (98).

If neither of the two mutual conditions just
mentioned holds then at the moment of reheating the condition (75) is not
valid, and the reheating temperature is determined by fermionic particle
production. In this latter case at the moment of reheating we have
$$
\Gamma_\chi^{\rm (res)} \ < \ \Gamma_\psi \, , \eqno (106)
$$
and in the condition (92) for reheating we should replace the value
$\Gamma_\chi^{\rm (res)}$ by $\Gamma_\psi$.
Then using expression (46) for $\Gamma_\psi$ we obtain for the value of the
inflaton field amplitude $\varphi_0$ at the moment of effective reheating the
well-known expression
$$
\varphi_0 \ \lsim \ 10^{- 2}\, f^2\, M_{\rm P} \, , \eqno (107)
$$
and for the reheating temperature we get
$$
T_{\rm rh} \ \sim \ 10^{- 2}\, \lambda^{1/4}\, f^2\, M_{\rm P} \, . \eqno (108)
$$

In the case (106) from the equation (74) we obtain the following expression
for $\Gamma_\chi^{\rm (res)}$
$$
\Gamma_\chi^{\rm (res)} \ \simeq \ {\Gamma_\psi \over 8\, \pi^2\, \rho}\,
\sum_{\rm all\ resonance\ bands} \omega_{\rm res}^4 \,
\sinh^2\left( {\pi |g_n|^2 \over 2\Gamma_\psi\,\omega_{\rm res}^3}
\right)\,  . \eqno (109)
$$

It is useful to compare this value of $\Gamma_\chi^{\rm (res)}$ to the value
of $\Gamma_\chi$
given by (47). To do this note that mainly two resonance bands
contribute to the sum (109), namely, those given by (77) and (78).
Therefore we can write
$$
\Gamma_\chi^{\rm (res)} \ = \ \Gamma^{\rm (res)}_1(\sigma) + \Gamma^{\rm
(res)}_2(h) \, , \eqno (110)
$$
where $\Gamma^{\rm (res)}_1$ depends only on the coupling $\sigma$, and
$\Gamma^{\rm (res)}_2$
only on the coupling $h$. These two terms
in the sum (110) correspond to the two terms in the expression (47) for
$\Gamma_\chi$. We then rewrite the expression for $\Gamma_\chi$ in the similar
way
$$
\Gamma_\chi \ = \ \Gamma_1(\sigma) + \Gamma_2(h) \, . \eqno (111)
$$
Now using the expressions (46), (47), (65), (77) and (78) we can easily
estimate the ratios of each of the two terms in (110) to its corresponding
term in (111) as
$$
{\Gamma^{\rm (res)}_n \over \Gamma_n} \ \simeq \ {1 \over x_n} \sinh^2\,x_n \,
, \eqno (112)
$$
where
$$
x_n \ = \ {\pi\, |g_n|^2 \over 2\Gamma_\psi\, \omega_{\rm res}^3}
\, . \eqno (113)
$$

We can see that parametric resonance effects for the scalar field $\chi$ would
be
significant in the case (106) if the values of $x_n$ given
by (113) were larger then unity. This last condition however can be
shown to be incompatible with the assumption that fermionic particle
production dominates at the moment of efficient reheating.
Hence at this moment the value $\Gamma_\chi$ given by
(47) is larger than the resonance value $\Gamma^{\rm (res)}_\chi$ given under
the conditions considered by (109).

All the results obtained in this section for the inflaton field amplitude
during effective reheating can be written in a single equation
$$
\varphi_0 \ \lsim \ M_{\rm P} \, \max\, \left\{10^{- 2} f^2, \
10^{- 1} \left({h \over \lambda}\right)^2\, \log^{- 1}\, {1 \over \lambda}, \
\left({\sigma \over M_{\rm P}\, \lambda} \right)^{2/3}\,
\log^{- 1/3}\, {1 \over \lambda} \right\} \, , \eqno  (114)
$$
from which also all the necessary inequalities between coupling
constants can be easily derived.
We stress once again that for the values of $\varphi_0$ given by
(114) the condition $4 H < \Gamma_\psi + \Gamma_\chi^{\rm (res)}$ is valid
so the reheating process is very effective.

The expression for the temperature of reheating is then
$$\eqalign{
& T_{\rm rh} \ \sim \ \lambda^{1/4}\, \varphi_0 \ \sim \cr
& M_{\rm P} \,
\lambda^{1/4}\, \max\, \left\{10^{- 2} f^2, \
10^{- 1} \left({h \over \lambda}\right)^2\, \log^{- 1}\, {1 \over \lambda},
\left({\sigma \over M_{\rm P}\, \lambda} \right)^{2/3}\,
\log^{- 1/3}\, {1 \over \lambda} \right\} \, . \cr} \eqno  (115)
$$

It is very instructive to compare our results (114), (115) to those which
would be obtained by making use of Born approximation (described in
Section 3)
for the scalar particles $\chi$ production rates. If instead of
$\Gamma_\chi^{\rm (res)}$ we used the
value $\Gamma_\chi$ given by (47) then we would arrive at the following
reheating conditions
$$
\varphi_0 \ \lsim \  M_{\rm P} \, \max\, \left\{10^{- 2} f^2\,\, , \hskip.5cm
10^{- 2} \lambda\,\left({h \over \lambda}\right)^2\,\, ,
\hskip.5cm \lambda\, \left({\sigma \over M_{\rm P}\, \lambda} \right)^{2/3}
\right\} \, , \eqno  (116)
$$
and
$$
T_{\rm rh} \ \sim \  M_{\rm P} \, \lambda^{1/4}\, \max\,
\left\{10^{- 2} f^2\,\, , \hskip.5cm
10^{- 2} \lambda\,\left({h \over \lambda}\right)^2\,\, ,
\hskip.5cm \lambda\, \left({\sigma \over M_{\rm P}\, \lambda} \right)^{2/3}
\right\} \, . \eqno  (117)
$$
One can see that for the scalar particle couplings the values (116), (117)
are smaller by an enormous factor $\left[\lambda\, \log(1/ \lambda)\right]
^{- 1}$ than the corresponding correct estimates (114) and (115).
\medskip
{\bf \chapter{Case of potential $V(\varphi) = {1 \over 2}\, m^2 \varphi^2$}}

In this section the analysis will procede along the
same lines as in the previous one. In the case of a quadratic inflaton
potential we get from (31), (32)
$$
\omega \ = \ m \, , \eqno (118)
$$
so the inflaton field oscillates with constant frequency. The resonance
frequency
$\omega_{\rm res} = n \omega / 2$ will thus be constant in time. For the time
derivative factor in (62) we then obtain, taking into account (58)
$$
\Big| {d\Delta\omega_k \over dt} \Big|_{\Delta\omega_k = 0} \ = \
H\, \omega_{\rm res} \, , \eqno (119)
$$
and from (64) we will have
$$
\Gamma_\chi^{\rm (res)} \ = \ {H \over 2\, \pi^2\, \rho}\, \sum_{\rm all\
resonance\ bands}\,
N_{\rm res}\, \omega_{\rm res}^4 \, , \eqno (120)
$$
where all the relevant variables have been explained in Section 4.
The expressions (59), (71) remain valid in the case considered but the
estimate analogous to (72) will be modified as follows
$$\eqalign{
\int \mu_+ dt \, &\approx \,  \int\limits_{- |g_n|/2\omega_{\rm
res}}^{|g_n|/2\omega_{\rm res}}
{1 \over 2 \omega_{\rm res}} \sqrt{|g_n|^2  -
\Delta\omega_k^2(2\omega_{\rm res})^2}\,\, {1 \over H\,
\omega_{\rm res}}\, d \Delta\omega_k \,  \approx \cr
& \approx \,\,
{|g_n|^2 \over 4\, H\,\omega_{\rm res}^3}
\, \int \limits_{-1}^1 \sqrt{1 - x^2}\, dx \, = \, {\pi \over 8} \,
{|g_n|^2 \over H\,\omega_{\rm res}^3}  \, , \cr} \eqno (121)
$$
so that the occupation numbers $N_{\rm res}$ will be given by
$$
N_{\rm res} \ \simeq \ \sinh^2 \left({\pi \over 8}\, {|g_n|^2 \over H\,
\omega_{\rm res}^3} \right)  \, , \eqno (122)
$$
and the equation (120) for $\Gamma_\chi^{\rm (res)}$ will read
$$
\Gamma_\chi^{\rm (res)} \ \simeq \ {H \over 2\, \pi^2\, \rho}\, \sum_{\rm all\
resonance\ bands}
\omega_{\rm res}^4 \,
\sinh^2 \left({\pi \over 8}\, {|g_n|^2 \over H\, \omega_{\rm res}^3}
\right) \,  . \eqno (123)
$$

We must compare this value to the value of $\Gamma_\chi$ obtained
in the Born approximation. Writing $\Gamma_\chi^{\rm (res)}$ and $\Gamma_\chi$
in the form (110) and
(111) and using the expressions (47), (77) and (78), we will arrive at an
estimate of the same form as (112) but with different $x_n$:
$$
x_n \ = \ {\pi \over 8}\, {|g_n|^2 \over H\, \omega_{\rm res}^3}
\, . \eqno (124)
$$
Again we see that parametric resonance effects for the scalar field $\chi$
are significant if the values $x_n$ are large.

As usual, one of the terms in the sum (123) will dominate. In the case
$\varphi_0 < \sigma/h$ the harmonic with $n = 1$ dominates, whereas for
$\varphi_0 > \sigma/h$ the harmonic corresponding to $n = 2$ is more important.
The relevant expressions for the amplitudes in these two cases are given by
(77) and (78).

In the case $\varphi_0 < \sigma/h$ we get
$$
\Gamma_\chi^{\rm (res)} \ \simeq \ {H\, m^2 \over 16\, \pi^2\, \varphi_0^2}
\sinh^2 \left(
{\pi\, \sigma^2\, \varphi_0^2 \over H\, m^3} \right)\, , \eqno (125)
$$
and the condition of effective particle production (92) in this case will read
$$
\left({\varphi_0 \over m}\right)^2 \ \gsim \ {H m \over \pi \sigma^2}\,
\log {H m \over \sigma^2} \, . \eqno (126)
$$
As soon as particle creation begins the Hubble parameter depends on
the total energy density of the inflaton field and of the particles created.
So the value of the Hubble parameter definitely exceeds the expression
given by (12) with only inflaton energy density taken into account. This
enables us to obtain the lowest possible bound on the region of the effective
particle production. If we substitute in (126) the lowest possible value for
the Hubble parameter given by (12) we will get the result, up to numerical
factor of order one,
$$
{\varphi_0 \over m} \ \gsim \ {m^3 \over \sigma^2\, M_{\rm P}} \, \log {m^3
\over
\sigma^2\, M_{\rm P}} \, . \eqno (127)
$$
This estimate is valid only if the coupling constant $\sigma$ is in the range
given by
$$
\left({m \over M_{\rm P}}\right)^2 \ \ll \left({\sigma \over m}\right)^2 \
\ll \ {m \over M_{\rm P}} \, . \eqno (128)
$$
The first inequality in (128) is the condition for perturbation theory to
be valid, it
stems from (44). The second inequality in (128) is the criterion of large
occupation numbers of the particles produced, that is, of
effectiveness of the resonance particle production (in other terms, it is the
condition for the value $x_1$ given by (124) to be larger then unity).
The validity of the adiabaticity condition (61) in this case follows from
(127) and from the first inequality in (128).

In any case there is also another region in which the condition (92) holds so
that particle production can be effective. This region is given by
$$
{\varphi_0 \over m} \ \lsim \ {\sigma^2\, M_{\rm P} \over m^3} \, , \eqno (129)
$$
and it can be obtained by using the first term in the expression (47) for
the value $\Gamma_\chi$ since parametric resonance in this region is not
effective.

In the case opposite to the one just considered, namely, when $\varphi_0 >
\sigma/h$ we obtain
$$
\Gamma_\chi^{\rm (res)} \ \simeq \ {H\, m^2 \over \pi^2\, \varphi_0^2}  \sinh^2
\left(
{ \pi\, h^2\, \varphi_0^4 \over 32\, H\, m^3} \right)\, , \eqno (130)
$$
and the condition (92) will give the estimate
$$
\left({\varphi_0 \over m}\right)^4 \ \gsim \ {32 H \over \pi h^2 m }\,
\log {H \over h^2 m} \, , \eqno (131)
$$
The lowest possible bound on the effective particle production region in
this case will be obtained if we substitute the lowest possible value for
$H$, given by (12), into the right-hand side. We will get
$$
\left({\varphi_0 \over m} \right)^3 \ \gsim \ {65 m \over h^2\, M_{\rm P}}\,
\log {m \over h^2\, M_{\rm P}} \, . \eqno (132)
$$
The theory which led to the result (132) will be self-consistent if the
coupling $h$ is in the range
$$
\left({m \over M_{\rm P}} \right)^4 \ \ll \ h^2 \ \ll {m \over M_{\rm P}} \, .
\eqno (133)
$$
Again, the first inequality in (133) is the requirement that perturbation
theory be valid
and the second one is the requirement of large occupation numbers of the
particles produced. For the values of $\varphi_0$ (132) the adiabaticity
condition (61) can be shown to hold.

We emphasize that the condition (92) in the case considered determines
not the possible reheating temperature but rather the ``freeze-out''
boundary (127) or (132) for
the inflaton field $\varphi$: below the values of $\varphi_0$ given by these
expressions the particle creation process is ineffective and the inflaton field
energy density decrease is dominated by the
cosmological redshift rather then by particle production. The energy
density
of ultrarelativistic particles created decreases more rapidly than that of the
inflaton field and the universe can well become again dominated by
the inflaton field $\varphi$ until the condition (129) is satisfied. If the
couplings $\sigma$ and $h$ are extremely small so that
$$
\left({\sigma \over m}\right)^2 , \ \ h \ \ \ \ll \ \ \ \left({m \over
M_{\rm P}}\right)^2 \, , \eqno (134)
$$
then the effective reheating condition (92) in the model with quadratic
inflaton potential takes place only for the values of $\varphi_0$ given by
(129). It is easy to see that in all the cases considered above the minimal
possible ``freeze-out'' energy density $\rho_{\rm freeze}$ of the inflaton
field is
$$
\rho_{\rm freeze}  \ \sim \ m^4 \, . \eqno (135)
$$

If during the particle creation process the energy density of the universe
becomes dominated by the particles produced, then the estimates for the
actual values of the inflaton field energy density at which it freezes out
will be higher then those given by (127) and (132). Such a condition is
likely to take place in the range of coupling constants given by (128)
and (133) since, as discussed in [\kls], at the onset of the oscillation
period of the inflaton, reheating process may be very efficient. In this
case one must use the equations (126) and (131)
in which it should be taken into account that the Hubble parameter $H$ is
dominated by the contribution from the particles created and, hence, is much
higher then given by Eq. (18). We note that in the range of couplings (128)
and (133), at the moment of the onset of the inflaton oscillations,
perturbation theory
developed in this paper breaks down. It becomes valid again when the
amplitude $\varphi_0$ decreases sufficiently and the conditions (44) start
to hold. The value of the Hubble parameter in the equations
(126) and (131), however, cannot be higher then its value $H \sim m$ at
the end of inflation. This allows one to estimate also the maximal possible
boundaries for the inflaton field freezing out which will be [\kls]
$$
\left({\varphi_0 \over m}\right)^2 \ \sim \ \left({m \over \sigma}\right)^2\,
\log {m \over \sigma} \, , \eqno (136)
$$
and
$$
\left({\varphi_0 \over m}\right)^4 \ \sim \ h^{- 2}\, \log h^{- 2}
\, . \eqno (137)
$$
These estimates are just on the boundary of the applicability of perturbation
theory.

If interactions of the inflaton $\varphi$ with fermions take
place, we obtain, using the equation (46), the following estimate
$$
\varphi_0  \ \lsim \ 10^{- 2} f^2\, M_{\rm P} \, \eqno (138)
$$
for the reheating condition, which coinsides with the analogous estimate (107)
in the case of the potential $V(\varphi) = \lambda \varphi^4$ which we
considered in the previous section. It can be shown, using the expression (31)
for the value $\omega$ and the expression (46) for the value of $\Gamma_\psi$,
that the estimate (138) does not depend on the power $q$ of the inflaton
potential $V(\varphi)$.
\medskip
{\bf \chapter{Discussion}}

In this paper we considered the problem of scalar particle production after
inflation by a quickly oscillating inflaton field. We were using the
framework of the
chaotic inflation scenario with quartic and quadratic inflaton potentials,
and we considered linear and quadratic coupling of the inflaton field $\varphi$
to a quantum scalar field $\chi$ (see formula (28) for the interaction
Lagrangian). The particle production effect has been studied
using the standard technique of Bogolyubov transformations.
Specific attention
has been paid to parametric resonance phenomena which take place in the
presence of a quickly oscillating inflaton field.

We have found that in the region of applicability of perturbation theory
(when inequalities (44) are valid) the
effects of parametric resonance are crucial, and estimates based on
first order Born approximation are often not correct.

In the case of the quartic
inflaton potential $V(\varphi) = \lambda \varphi^4$ the reheating process is
very efficient for any type of
coupling between the inflaton field and the scalar field $\chi$ even for
small values of coupling constants. The reheating temperature in this case is
$\left[\lambda\, \log\, (1 / \lambda)\right]^{- 1}$ times larger than the
corresponding estimates based on first order Born approximation (see the
expressions (115), (117)).

In the case of a quadratic inflaton potential the situation is more
complicated.
The reheating process
depends crucially on the type of coupling between the inflaton and the scalar
field $\chi$ and on the magnitudes of the coupling constants $\sigma$ and $h$.
The theory predicts not ony the possible reheating temperature but also the
effective energy density of the inflaton field ``freezing out''
below which the inflaton field energy density decrease is dominated by the
cosmological redshift rather then by particle production. The inflaton,
so to say, decouples from the matter it creates. In the case of the quadratic
inflaton potential we obtained the lowest (equations (127) and (132)) and
highest (equations (136) and (137)) possible
boundaries   of the effective energy density of the
inflaton field when it freezes out. With an interaction
linear in the inflaton field, besides these possible freeze-out
boundaries which exist for
couplings in the range (128), there is also a reheating boundary (129). For an
interaction quadratic in the inflaton field the lowest possible freeze-out
boundary (132)
exists in the range (133) of the values of coupling $h$. If the couplings
$\sigma$ and $h$ are as small as to satisfy (134) then the effective reheating
condition $\Gamma \gsim 4 H$ is achieved only at the later stages after
the condition (129) starts to hold.

The situation is more complex for more complicated potentials. For example,
one may wish to consider the inflaton potential of shape
$$
V(\varphi) \ = \ \lambda \left(\varphi^2 - \eta^2 \right)^2 \, , \eqno (139)
$$
with spontaneous symmetry breaking at a scale $\eta$. In the chaotic inflation
scenario the inflaton field rolls from large values of $|\varphi|$ to the
minimum of its potential. After inflation it starts quasiperiodic motion.
For values of the amplitude $\varphi_0 \gg \eta$ the behaviour of the
inflaton is like for a $\lambda \varphi^4$ potential, and reheating proceeds
as described in Section 5. But as soon as the magnitude of
$\varphi_0$ becomes close to $\eta$ the dynamics changes and the inflaton
field starts oscillating around one of the minima of its potential at
$|\varphi| =
\eta$. The inflaton potential in the vicinity of its minimum can be
approximated as quadratic, hence the reheating (freeze-out) will be described
by the theory of Section 6.

In parallel to our own work, Kofman, Linde and Starobinsky have also been
working on these issues for several years. Their results were recently
summarized in Ref. \kls. While our paper
deals only with the case when the conditions (44) hold and perturbation theory
in the inflaton couplings to the other fields is valid, both the cases (44)
and opposite to (44) are considered in [\kls]. The resonance frequency
bandwidth can be shown to be relatively narrow in the case (44), and broad
in the case opposite to (44), so these two cases are called in [\kls]
``narrow'' and ``broad'' resonance cases respectively. In the case of broad
resonance, higher resonance bands dominate the
contribution to the particle production. After the inflaton field amplitude
decreases sufficiently one enters into the regime of narrow resonance which
can be accurately
described by perturbation theory. The results announced in [\kls] for the case
of narrow resonance in general agree with those obtained
in our paper in Chapters 5 and 6. There are some results for the case of broad
resonance which we would like to compare with our results for the
narrow resonance case. In doing this we shall use the notation of the
present paper which somewhat differs from that of [\kls].

In the case of quartic inflaton potential the estimate of [\kls] $N_k(t)
\propto \exp\left({1 \over 5} \sqrt \lambda \varphi_0 t \right)$ for the
case of coupling $h = 6 \lambda$ (self-excitation of the inflaton, considered
in [\kls]) can be compared with our estimate $N_k(t) \propto
\exp(2 \mu_+ t) \simeq \exp\left({h \over 2 \lambda} \sqrt \lambda
\varphi_0 t \right)$ for the case of smaller coupling
($h \ll \lambda$).
We note that these estimates will be numerically of the same order of
magnitude if we choose our smaller coupling to be $h \simeq 2 \lambda / 5$.
For the values of $h \simeq
0.22 \lambda$ our estimate (93) for the beginning of reheating will be of the
same order
as the estimate of [\kls] $\varphi_0 \lsim 0.005 M_{\rm P}\, \log^{- 1}
(1 / \lambda)$ for $h = 6 \lambda$. These estimates show that the effect of
other particle production for the values of coupling about $h \gsim 0.3
\lambda$ may be larger than the effect of inflaton self-excitation.
This value of $h$ is on the border of the applicability of
perturbation theory. We cannot say anything about the reheating with couplings
$h \gsim \lambda$ as our perturbation theory breaks down in this case.

If the inflaton potential is quadratic then, as previously discussed by Kofman,
Linde and Starobinsky (see Ref. \kls), in the case of absence of inflaton
coupling to fermions
and without linear (in inflaton field) coupling to scalar particles, the
inflaton field will eventually decouple from the rest of the
matter, and the residual inflaton  oscillations may provide the (cold) dark
matter of the universe. Decoupling of the inflaton field occurs somewhere
between the values given by the right-hand sides of (132) (lowest possible
freeze-out boundary) and (137) (highest possible freeze-out boundary).

\endpage
{\bf Acknowledgments}

It is our pleasure to thank A.Dolgov, J.Frieman, L.Kofman, A.Linde,
A.Starobinsky and I.Tkachev
for valuable discussions. We also thank L.Kofman, A.Linde and A.Starobinsky
for pointing out mistakes in the first version of this paper.
This work started during the 1990 US-USSR Young
Cosmologists program. We are grateful to the US National Academy of
Sciences and the USSR Academy of Sciences for initiating and supporting
this program. The work of Y.S. was also supported by the National
Research Council under the COBASE program and that of J.T. in part by NSF
grant NSF-THY-8714-684-A01.

\endpage

{\bf Appendix A}

In this appendix we present necessary expressions which describe the effect
of parametric resonance in the lowest resonance band.

Consider the equation for the function $Y(t)$ of the following form
$$
\ddot Y + \left( \omega_0^2 + \epsilon g(\omega t) \right) Y \ = \ 0
\, , \eqno ({\rm A}1)
$$
where $\omega_0$ and $\omega$ are constant parameters, $g(x)$ is a
$2\pi$-periodic function, and $\epsilon$ is a small number. Our task
is to find the approximate solution to the equation (A1) in the first
instability band of the frequencies $\omega_0$. To do this we will use
Bogolyubov method of averaging \Ref\bm{N.Bogoliubov and Y.Mitropolsky
{\it Asymptotic Methods in the Theory of Non-Linear Oscillations}
(translated from Russian, Hindustan
Publishing Corpn., Delhi, 1961).} which implies perturbation
theory of certain type in the small parameter $\epsilon$.

The function $g(x)$ being periodic, it can be developed in Fourier series
as follows
$$
g(x) \ = \ \sum_{n = - \infty}^{\infty}\, g_n\, e^{i n x} \, , \eqno ({\rm A}2)
$$
with the amplitudes $g_n$ satisfying
$$
g_n^* \ = \ g_{- n} \, . \eqno ({\rm A}3)
$$
With no loss of generality (for a small parameter $\epsilon$) we can put
$g_0 = 0$ (redefining $\omega_0$ if necessary). For the following purposes
it will be convenient to introduce the phases $\alpha_n$ as follows
$$
g_n \ = \ |g_n|\, e^{i \alpha_n} \, . \eqno ({\rm A}4)
$$

Let us introduce the value $\Delta$ by the following relation
$$
\omega_0^2 \ = \ \left({p \over q}\, \omega \right)^2 + \epsilon \Delta
\, , \eqno ({\rm A}5)
$$
where $p / q$ is a rational non-contractible number ($p$ and $q$ are integers).
The equation (A1) then becomes
$$
\ddot Y + \left({p \over q}\, \omega \right)^2 Y \ = \ - \, \epsilon \left(
g(\omega t) + \Delta \right) Y \, . \eqno ({\rm A}6)
$$

The Bogolyubov method [\bm] consists in looking for a solution to the
equation
(A6) in the form of the following expansion in powers of $\epsilon$
$$
Y \ = \ a \cos \psi + \sum_{s = 1}^\infty \epsilon^s\,
u^{(s)}\left(a,\, \theta,\, {\omega \over q}t \right)\, , \eqno ({\rm A}7)
$$
where
$$
\psi \ = \ {p \over q}\, \omega t + \theta \, , \eqno ({\rm A}8)
$$
the $u^{(s)}$ are functions periodic in their second and third arguments,
$a$ is the amplitude
and $\theta$ is the phase of the solution (A7). The values of $a$ and
$\theta$ are not constant in the Bogolyubov approach. Rather, they are
functions of time $t$.

We will be concerned explicitly only with the first order approximation in
the small parameter $\epsilon$. For the function $u^{(1)}$ we have
$$
u^{(1)}\left(a,\, \theta,\, {\omega \over q} t \right) \ = \ \sum_{n = -
\infty}
^{\infty}\, u_n(a, \, \theta)\, e^{in{\omega \over q}t} \, , \eqno ({\rm A}9)
$$
with coefficients $u_n(a,\, \theta)$ periodic in $\theta$.

Clearly, in the case of $\epsilon = 0$ the solution to (A6) will be just the
first term in the right-hand side of (A7), with arbitrary constant amplitude
$a$ and phase $\theta$, so in this case we have $\dot a = 0$, $\dot \theta
= 0$. For $\epsilon \neq 0$ according to the Bogolyubov method we regard the
values
$\dot a$ and $\dot \theta$ as functions of $a$ and $\theta$ and we expand these
functions in powers of $\epsilon$ $$\eqalign{
& \dot a \ = \ \epsilon\, A(a,\, \theta) + {\cal O}\left(\epsilon^2\right) \, ,
\cr
& \dot \theta \ = \ \epsilon\, B(a,\, \theta) + {\cal O}\left(\epsilon^2\right)
\, . \cr} \eqno ({\rm A}10)
$$
Putting the anzatz (A7) into the equation (A6) and using (A10), after
collecting terms of first order in $\epsilon$ we will
get the following equation for the function $u^{(1)}$
$$
\ddot u^{(1)} + \left({p \over q} \omega \right)^2 u^{(1)} \ = \ -\, (g +
\Delta)\,
a \cos \psi + 2\, {p \over q}\, \omega A \sin \psi + 2\,
{p \over q}\, \omega
B\, a \cos \psi \, . \eqno ({\rm A}11)
$$

Using the expansions (A2) and (A9) we can develop both sides of the
equation (A11) in harmonics $\exp (i{n \over q} \omega t )$. On the
left-hand side of (A11) we will have, to the zeroth order in $\epsilon$,
$$
\sum_{n = - \infty}^{\infty} \Big[\left({p \over q}\right)^2 -
\left({n \over q}\right)^2 \Big]\, \omega^2 u_n\, e^{i {n \over q} \omega t}
\, , \eqno ({\rm A}12)
$$
and we see that terms with $n = \pm p$ in this expansion are equal to zero.
The corresponding terms in the development of the right-hand side of (A11)
should also vanish. These conditions allow us to determine the functions
$A(a,\, \theta)$ and $B(a, \, \theta)$. If we denote by $s$ the number
$$
s \ = \ {2 p \over q} \, , \eqno ({\rm A}13)
$$
then the expressions for these functions will look as follows
$$\eqalign{
& A(a,\, \theta) \ = \ {a |g_s| \over s \omega} \sin(2 \theta - \alpha_s)
\, , \cr
& \cr
& B(a,\, \theta) \ = \ {1 \over s \omega} \left( \Delta + |g_s| \cos(2 \theta
- \alpha_s) \right) \, , \cr} \eqno ({\rm A}14)
$$
with the phases $\alpha_s$ defined in (A4) for $s$ integer. The expressions
(A14) are valid for arbitrary $s$, in the case of $s$ non-integer we should
simply put $g_s$ to zero, and the (unspecified) value of $\alpha_s$ is of no
importance.

If we make the change of variables
$$\eqalign{
& x \ = \ a \cos\left(\theta - {\alpha_s \over 2} \right) \, , \cr
& \cr
& y \ = \ a \sin\left(\theta - {\alpha_s \over 2} \right) \, , \cr}
\eqno ({\rm A}15)
$$
then the system (A10) in terms of new variables $x$ and $y$ will look as
follows
$$\eqalign{
& \dot x \ = \ {\epsilon \over s \omega}\, (|g_s| - \Delta)\, y +
{\cal O}\left(\epsilon^2\right) \, , \cr
& \cr
 & \dot y \ = \ {\epsilon \over s \omega}\, (|g_s| + \Delta)\, x +
{\cal O}\left(\epsilon^2\right) \, . \cr}
\, \eqno ({\rm A}16)
$$

To first order in $\epsilon$ the system (A16) is linear and its eigenvalues
are
$$
\mu_{\pm} \ = \ \pm\, {\epsilon \over s \omega}\, \sqrt{|g_s|^2 - \Delta^2}
\, . \eqno ({\rm A}17)
$$
  From this expression one can see that $\mu_\pm$ can be real only if
$g_s \neq 0$, that is if $s$ is integer. Hence unstable growth of the
solution $Y$ to the equation (A6) can take place only for the values of
$\Delta$ ($\Delta$ is defined in (A5))
$$
\Delta \ < \ |g_s| \, , \eqno ({\rm A}18)
$$
and the growth (decay) rate is exponential, $Y \sim exp\, (\mu_\pm t)$,
with $\mu_\pm$ given by (A17).

For every integer $s$ for which $g_s$ is
non-zero the expression (A17) gives the growth rate of the
solution $Y$ in the first instability band determined by
(A5) and (A18). There are also infinite number of instability bands whose
width and growth rates are higher order in $\epsilon$ (to all orders of
$\epsilon$). Relevant expressions can be obtained to any desirable
order in $\epsilon$ by keeping track of the higher order terms in the
expansions (A7) and (A10). For the purpose of this paper these higher order
expressions are not required.
\medskip
{\bf Appendix B}

The aim of this Appendix is to provide necessary detailes of the resonance
particle
production effect studied in Section 3 of this paper, in particular, to
derive the formula (60). We first consider the general problem of a harmonic
oscillator with time-dependent frequency, and after that the problem of
resonant scalar particle production.
\medskip

{\bf a) Harmonic oscillator with time-dependent frequency.}

Consider an oscillator with coordinate $Q$, conjugate momentum $P$, whose
frequency $\Omega$ depends on time so that the Hamiltonian is
$$
{\cal H} \ = \ {1 \over 2}\, \left(P^2 + \Omega^2(t)\, Q^2\right)
\, . \eqno ({\rm B}1)
$$

For a quantum oscillator the values $Q$ and $P$ are Hermitian operators with
standard commutation relations (we put $\hbar = 1$)
$$
[Q,\, P] \ = \ i \, . \eqno ({\rm B}2)
$$
In the Schr\"odinger representation, these operators are time-independent.

Our aim will be to go to a convenient time-dependent frame in
Hilbert space, in which the Hamiltonian (B1) is diagonal at every moment of
time. To do this define
time-dependent operators $a(t)$ and $a^\dagger (t)$ by
$$\eqalign{
a \ \ & = \ {e^{i\int \Omega dt} \over \sqrt{2\Omega}} \left(\Omega Q + iP
\right) \, , \cr
& \cr
a^\dagger \ & = \  {e^{- i\int \Omega dt} \over \sqrt{2\Omega}}
\left(\Omega Q - iP \right) \, . \cr} \, \eqno ({\rm B}3)
$$
These operators are mutually Hermitian conjugate and have the standard
commutation relations of creation-annihilation operators
$$
[a,\, a^\dagger] \ = \ 1 \, . \eqno ({\rm B}4)
$$
In terms of these operators the Hamiltonian is expressed as follows
$$
{\cal H} \ = \ \Omega \left( {1 \over 2} + a^\dagger a \right) \, . \eqno ({\rm
B}5)
$$

The orthonormal frame in Hilbert space which diagonalizes the Hamiltonian
${\cal H}$ is given by time-dependent states
$$
|n_t> \ = \ {\left(a^\dagger (t)\right)^n \over \sqrt{n!}}\, |0_t>
\, , \eqno ({\rm B}6)
$$
constructed from the time-dependent vacuum state $|0_t>$ which is annihilated
by the operator $a(t)$.

The operators $a(t)$ and $a^\dagger(t)$ obey the following equations of motion
$$\eqalign{
\dot a \ \ & = \ {\dot \Omega \over 2 \Omega}\, e^{2i\int \Omega dt}\,
a^\dagger \, , \cr
& \cr
\dot a^\dagger \ & = \ {\dot \Omega \over 2 \Omega}\, e^{- 2i\int \Omega dt}\,
a\, , \cr } \, \eqno ({\rm B}7)
$$
so these operators vary with time only when $\Omega$ varies \footnote{1)}
{This property is achieved by a convenient choice of the time-dependent
phase factors in the definition (B3) of the operators $a$ and $a^\dagger$.}.

The solution to the equations (B7) can be written in terms of constant
creation and annihilation operators $a_0^\dagger$ and $a_0$ as follows
$$\eqalign{
a(t) \ \ & = \ \alpha(t)\, a_0 + \beta^*(t)\, a_0^\dagger \, , \cr
a^\dagger (t) \ & = \ \beta(t)\, a_0 + \alpha^*(t)\, a_0^\dagger \, , \cr}
\eqno ({\rm B}8)
$$
with the complex functions $\alpha(t)$ and $\beta(t)$ obeying the system
of equations similar to (B7)
$$\eqalign{
\dot \alpha \ & = \ {\dot \Omega \over 2 \Omega}\, e^{2i\int \Omega dt}\,
\beta \, , \cr
& \cr
\dot \beta \ & = \ {\dot \Omega \over 2 \Omega}\, e^{- 2i\int \Omega dt}\,
\alpha\, . \cr } \, \eqno ({\rm B}9)
$$
The system (B8) represents what is called Bogolyubov transformation between
two pairs of creation - annihilation operators.

If the oscillator initially (at $t = 0$) is in the vacuum state then its
state $|0_0>$ is annihilated by the operator $a_0$ and the initial conditions
for the functions $\alpha(t)$ and $\beta(t)$ are
$$
|\alpha(0)| \ = \ 1 \, \, , \hskip.3cm \beta(0) \ = \ 0 \, \, . \eqno ({\rm
B}10)
$$
At the moment $t$ it will not be in the vacuum state $|0_t>$, annihilated by
the operator $a(t)$. Rather, there is the following relation between the states
considered
$$
|0_0> \ = \ {1 \over \sqrt{|\alpha(t)|}}\,\, \exp \left( {\beta^*(t)
\over 2
\alpha^*(t)} \left(a^\dagger(t)\right)^2\right)\, |0_t> \, , \eqno ({\rm B}11)
$$
which follows from the equations (B8). At the moment $t$ the average number of
the oscillator excitation level (number of quanta) is
$$
N(t) \ = \ <0_0|\, a^\dagger (t)\, a(t)\, |0_0> \ = \ |\beta(t)|^2
\, . \eqno ({\rm B}12)
$$

In the Heisenberg representation, the equation of motion for the operator
$Q(t)$ is
$$
\ddot Q \ + \ \Omega^2\, Q \ = \ 0 \, . \eqno ({\rm B}13)
$$
The solution to this equation can be expressed in terms of the operators
$a_0$ and $a_0^\dagger$ so that
$$
Q(t) \ = \ Q^{(-)}(t)\, a_0 + Q^{(+)}(t)\, a_0^\dagger \, . \eqno ({\rm B}14)
$$
Comparing this expression with the equations (B3) and (B8) we can express the
coefficient
$\beta(t)$ in terms of $Q^{(-)}(t)$ at the moment of time at which the first
time derivative of the
frequency $\Omega$ vanishes as follows
$$
\beta \ = \ {1 \over \sqrt{2 \Omega}} e^{- i \int \Omega dt} \left(
\Omega Q^{(-)} - i\dot Q^{(-)} \right) \, . \eqno ({\rm B}15)
$$
\medskip

{\bf b) Resonance particle production.}

Now let us turn to the issue of scalar particle creation in the external
spatially
homogeneous periodic field $\epsilon g(\omega t)$ whose properties were
described in Appendix A (see the equations (A2)-(A4)). Remember that
$\epsilon$ is a small parameter of the perturbation theory. The scalar
field operator $\chi$ which describes particles with
mass $m_\chi$ can be decomposed into spatial Fourier modes
$$
\chi \ = \ \int {d^3k \over (2\pi)^3}\, Q_k\, e^{ikx}
\, , \eqno ({\rm B}16)
$$
so that every mode corresponds to a quantum oscillator with Hamiltonian (B1),
complex coordinate $Q = Q_k$ and frequency $\Omega = \Omega_k$ given by
$$
\Omega^2_k \ = \ \omega^2_k  + \epsilon g(\omega t)
\, , \eqno ({\rm B}17)
$$
where $\omega_k^2 = m_\chi^2 + k^2$. Because the frequency $\Omega_k$
depends on time, the oscillator with label $k$ will be excited and this
means particle production. We can use all the above
expressions obtained for the generic oscillator by simply adding a
subscript $k$. Thus, the theory of scalar particle production
in the case under consideration is only a slight
modification of the oscillator formalism described above. One only has to
take into account correctly the presence
of infinite number of oscillators labelled by wavenumbers $k$ and
interrelations between them.
The expression analogous to (B11) relating the initial scalar field vacuum
to the vacuum
at the moment of time $t$ will look like follows
$$
|0_0> \ = \ \prod_k \, {1 \over \sqrt{|\alpha_k(t)|}}\,\, \exp \left(
{\beta_k^*(t) \over 2\alpha_k^*(t)}\, a^\dagger_k(t)\, a^\dagger_{- k} (t)
\right)\, |0_t> \, , \eqno ({\rm B}18)
$$
where $a^\dagger_k (t)$ and $a_k(t)$ are the time-dependent creation and
annihilation operators which diagonalize the scalar field Hamiltonian at the
moment $t$. From the expression (B18) it can be seen that scalar particles
are created in pairs
with opposite wavenumbers $k$. The average number of particles produced
at the moment $t$ in the $k$-th mode will be given by the expression similar
to (B12) with the only modification being the addition of subscript $k$. To
determine the average
number $N_k$ of particles produced we can use the formulas (B12) and (B15).
We only need to know the solution for $Q_k^{(-)}(t)$. This function obeys the
equation
$$
\ddot Q_k^{(-)} \ + \ \left( \omega_k^2 + \epsilon g(\omega t) \right)
Q_k^{(-)}
\ = \ 0 \, , \eqno ({\rm B}19)
$$
which is just the equation (A1) which we considered in the Appendix A. Initial
conditions for $Q_k^{(-)}$ stem from the initial vacuum conditions (B10) and
are
$$
|Q_k^{(-)}(0)| \ = \ {1 \over \sqrt{2 \Omega_k(0)}} \, , \hskip1cm
\left(\Omega_k Q_k^{(-)} - i\dot Q_k^{(-)} \right)\Big|_{t = 0} \ = \ 0
\, . \eqno ({\rm B}20)
$$

The solutions to the equation (B19) in the resonance band were studied in the
Appendix A. According to (A5) we introduce the value $\Delta$ by
$$
\omega_k^2 \ = \ \left({s \over 2}\, \omega \right)^2 + \epsilon \Delta
\, , \eqno ({\rm B}21)
$$
where we replaced $p/q$ in advance by the half-integer $s/2$ according to the
results of the
Appendix A. Now the solution to $Q_k^{(-)}$ will be given by a complex linear
combination of the solutions of type (A7), namely
$$
Q_k^{(-)} \ = \ {1 \over \sqrt{2\omega_k}} \left(a_1 \cos \psi_1 + ia_2\,
 \cos \psi_2\right) + {\cal O}(\epsilon) \, , \eqno ({\rm B}22)
$$
with
$$
\psi_{1,2} \ = \ {s \over 2}\, \omega t + \theta_{1,2} \, . \eqno ({\rm B}23)
$$
The initial conditions for $a_{1,2}$ and $\theta_{1,2}$ stem from (B20)
$$
a_1 \ = \ a_2 \ = \ 1 \, , \hskip1cm \theta_2 - \theta_1 \ = \ {\pi \over 2}
\, . \eqno ({\rm B}24)
$$

If in analogy to (A15) we introduce the variables $x_{1,2}$ and $y_{1,2}$ as
\footnote{1)}{The constant value of the phase $\alpha_n$ defined in (A4)
can be eliminated without loss of generality by a shift of the values of
$\theta_{1,2}$.}
$$
\eqalign{
x_1 \ = \ a_1 \cos \theta_1\, ,\hskip2cm
& x_2 \ = \ a_2 \cos \theta_2\, , \cr
y_1 \ = \ a_1 \sin \theta_1\, , \hskip2cm
& y_2 \ = \ a_2 \sin \theta_2\, , \cr} \eqno ({\rm B}25)
$$
then to first order in the small parameter $\epsilon$ the equations for
$x_{1,2}$ and $y_{1,2}$ will be identical to (A16). The initial conditions for
these functions follow from (B24)
$$\eqalign{
& x_1(0) \ = \ y_2(0) \, , \hskip.5cm x_2(0) \ = \ - \, y_1(0) \, , \cr
& |x_1|^2 + |y_1|^2 \ = \ |x_2|^2 + |y_2|^2 \ = \ 1\, . \cr } \eqno ({\rm B}26)
$$

To the lowest order in $\epsilon$, the expression for the value
$\beta_k$ is given by
$$
\beta_k \ = \ {1 \over 2} \left(x_1 + ix_2 + i(y_1 + iy_2) \right)
\, . \eqno ({\rm B}27)
$$
It obeys the equation
$$
\ddot \beta_k \ = \ \mu_+^2\, \beta_k \, ,  \eqno ({\rm B}28)
$$
which follows from the system (A16), with $\mu_+$ given by (A17).
The initial conditions for the value $\beta_k$ stem from (B26) and are
$$
\beta_k(0) \ = \ 0 \, \hskip2cm |\dot \beta_k(0)| \ = \ \mu_+|_{\Delta
= 0} \, . \eqno ({\rm B}29)
$$
Solving the linear system (B28) for the function $\beta_k$ with
the initial conditions (B29)  we
obtain the expression for the mean number of the particles produced
$$
N_k \ \equiv \ |\beta_k|^2 \ = \ {1 \over 1 - \Delta^2 / |g_s|^2}
\,\,  \sinh^2\left(\mu_+t\right) \, , \eqno ({\rm B}30)
$$
where as we recall, the value $\mu_+$ is given by (A17).

The analysis developed in this Appendix can be also used in the case when
the value of $\mu_+$ is not constant in time but changes adiabatically.
Namely, if
$$
|\dot \mu_+ | \ \ll \ \mu_+^2 \, , \eqno ({\rm B}31)
$$
then the mean number of the particles created can be approximated by
$$
N_k \ \simeq \ \sinh^2 \left(\int \mu_+\, dt \right)
\, , \eqno ({\rm B}32)
$$
where the integral in the argument of the hyperbolic sine is taken over
the time during which resonance condition (A18) is satisfied for the mode
with wavenumber $k$, and the prefactor similar to that of (B30) has been set
to unity which is reasonable because at the edge of the resonance waveband
(when $\Delta \approx |g_s|$)
the adiabaticity condition (B31) ceases to be valid.
\endpage

\refout
\par
\end